\documentclass[aps, notitlepage, nofootinbib]{revtex4-1}

\usepackage{geometry}
\usepackage{epsfig}
\usepackage{epstopdf}
\usepackage{bm}

\begin{document}

\title{Quantum Theory without Quantization}
\author{Ken Wharton}
\affiliation{Department of Physics and Astronomy, San Jos\'{e} State University, San Jos\'{e}, CA 95192-0106}
\begin{abstract}

The only evidence we have for a discrete reality comes from quantum measurements; without invoking these measurements, quantum theory describes continuous entities.  This seeming contradiction can be resolved via analysis that treats measurements as boundary constraints.  It is well-known that boundaries can induce apparently-discrete behavior in continuous systems, and strong analogies can be drawn to the case of quantum measurement.  If quantum discreteness arises in this manner, this would not only indicate an analog reality, but would also offer a solution to the so-called ``measurement problem''.

\end{abstract}

\maketitle

\setlength{\baselineskip}{1.2\baselineskip} 

\section{Introduction}

Even if we were in possession of a unified physical theory that explained all known experimental and observational data, we would still not be able to definitively answer the question of whether reality was fundamentally continuous or discrete.  If our unified theory was based on a continuous ontology, there would always be the possibility that this was simply an approximation for an underlying discrete reality, too fine for our experiments to detect.  Conversely, if the unified theory were based on discrete elements, future experiments might one day reveal that those elements emerged from an underlying continuous structure.  Examples in this category include solitons (particle-like solutions of non-linear wave equations), or the discrete modes in continuous boundary-value problems, such as laser cavities.

Of course, we are not in possession of a unified theory.  Of the two pillars of modern physics, general relativity (GR) is clearly in the continuum camp, but GR fundamentally conflicts with quantum theory (QT), a formalism with both continuous elements (quantum fields) and discrete elements (atomic energy levels, etc.).  Na\"ively, one might give more weight to GR's principle-based framework than to the wishy-washy position of QT.  But this view is not currently popular.  Many physicists think that QT implies a discrete substructure for GR's continuous spacetime, and the vast majority of the effort in reconciling these pillars lies in ``quantizing gravity'', as opposed to finding a GR-compatible revision of QT.

But this conclusion is by no means definitive, especially when one considers that one of the biggest unresolved issues of QT -- the ``Measurement Problem'' -- is directly related to this discrete/continuous division.  In a simplified nutshell, the QT formalism distinguishes between observations made from outside a system (which yield discrete results) and interactions within a system (which are treated much more continuously).  This isn't tenable because there's no objective definition of a ``system''; it can always be expanded to include any ``outside'' observations.  The measurement problem is how objectively different mathematical procedures could possibly be associated with an apparently subjective parsing of joint vs. separate systems.    

In this essay, I argue that the details of this unresolved tension (between the continuous and discrete aspects of QT) provide us with important clues as to whether nature is fundamentally digital or analog.  Far from implying that fundamentally nature is somehow \emph{both} (a meaningless position that many otherwise-rational people have tried to adopt), QT is more likely describing a parameter regime where one of these descriptions is emerging from the other.  After all, there are many known examples of discreteness arising from an underlying continuity, and vice-versa.  This essay argues that the best analogies to QT are found in the discrete modes of continuous boundary value problems.  If the discreteness of QT does indeed arise in this manner, then (together with GR) all of modern physics would be built upon a continuous foundation.  \footnote{Whether this foundation would be fundamental, or emergent from a deeper discreteness, would remain to be seen.}

\section{Discreteness without Measurements?}

Although discreteness only enters QT when one takes a measurement, many popular descriptions give the contrary impression that QT describes discrete entities, regardless of whether or not a ``measurement'' occurs.  This is a fault that lies partially with poor word choices (including the very word ``quantum''), but also arises from the natural assumption that measurements merely reveal an underlying reality.  Here is a typical quote from Wikipedia \cite{wiki}:  \begin{quote}A quantum mechanical system or particle that is bound -- that is, confined spatially-- can only take on certain discrete values of energy. This contrasts with classical particles, which can have any energy.  \end{quote}  

This quote implies that a quantum system must always exist in one of several discrete modes, but such a statement is contradicted by the mathematics of QT.  The fact is that a quantum system can be in any ``superposition'' of these different energy states, corresponding to a smooth continuum.  An unmeasured single electron in an atom is therefore completely unconstrained by anything discrete, a stubborn fact that is difficult for many students to wrap their heads around after learning about atomic shells and quantum numbers.

Of course, when an energy measurement is made on an atomic electron, discreteness finally rears its head; the electron is never found in a superposition of different energy states, but always has one of several discrete energy values.  One could argue that the mathematical structure of quantum mechanics is built around these discrete energies, but the way that these values come about is by solving a continuous boundary value problem (the wavefunction of the electron has to fall off to zero in all directions, sufficiently fast), making this an example of how quantized values naturally emerge when solving such problems in a bounded continuum.  (More examples of ``boundary-induced quantization'' will be covered in Section III.)

Retreating to something that doesn't have a spacetime representation -- like quantum spin -- doesn't help the case for a fundamental discreteness.   Inaccurate comments such as: `The spin of the electron can only be up or down' are simply false, unless one is talking about the result of an actual measurement.  In QT an electron spin lives on a continuous mathematical surface known as the Bloch sphere.  Indeed, there exists a well-known, one-to-one correspondence between the mathematics of an unmeasured (quantum) electron spin and the (classical) polarization of a plane electromagnetic wave.\cite{HnS,Klyshko}  This connection, combined with the clear continuum nature of the latter, makes it untenable to claim that former is built upon a discrete foundation.

The final line of retreat would be that quantum mechanics talks about discrete \emph{particles}, which is true to a point.  Quantum mechanics was \emph{designed} to discuss a well-defined number of particles; in that sense, a discreteness has been put in by hand.  But a more general theory which can handle particle creation and destruction -- quantum field theory (QFT) -- allows superpositions of different particle number states.  (This should be no surprise, as QFT is built upon continuous fields!)  Even when it comes to something as basic as particle number, nature is only discrete when we actually take a measurement.

But if things are discrete whenever we look, why wouldn't they also be discrete when we \emph{don't} look?  While this classical logic is tempting, it simply doesn't stand up to the most basic quantum experiments.  Take the double-slit experiment, where single particles pass through two closely-spaced slits (one particle at a time), and form a two-slit interference pattern as if the particles were continuous waves.  Whenever one looks to see which slit the particle passes through, the interference pattern disappears.  It only reappears when we \emph{don't} look.  Any theory which demanded the same behavior, whether there was a measurement or not, would be unable to explain a vast array of similar observations.  

Further insights can be gleaned from imprecise, or ``weak'' measurements.  True, all measurements are imprecise to some extent.  But there's evidence that this non-exactness is not merely due to measurement imprecision.  For example, every observed transition between two atomic energy states has a ``linewidth'', or range of possible energies, a non-discrete phenomenon that significantly impacts the real-world behavior of lasers.  In another example, if one sends particles through a slit, the resulting behavior is only explicable in terms a wave that passes through the \emph{entire span} of the slit.  As one changes the size of the slit, the discreteness somehow automatically compensates.  Dialing down the measurement sensitivity, as in recent ``weak measurement'' experiments, has also been observed to gradually destroy the discreteness predicted by quantum mechanics.\cite{AAV,WeakExp}  It is as if particles somehow know how precisely they are being measured, and conspire to only be as discrete as the measurement precision demands. 

This behavior is consistent with the known limit: If one takes away all measurements, QT turns out to be not quantized at all.  This is in itself evidence of a continuous foundation for QT -- with the enormous elephant in the room being the measurement problem, and the apparent discreteness that this process somehow generates.  In the next sections, I'll talk more about how the act of `looking' could have such dramatic consequences, and how this might be our biggest clue as to what is really going on.

\section{Boundary-Induced Quantization}

It is well known that constraining a continuous system with external boundaries can lead to an emergent discreteness.  The most widely familiar example, often seen in science museums, is a vibrating string -- when fixed at both ends, the string preferentially vibrates at frequencies which form a standing wave pattern with an integer number of half-wavelengths between the two boundaries.  I'll call this emergence of discrete modes Boundary-Induced Quantization, or BIQ for short. 

The same phenomenon occurs in laser cavities, where the resonant frequencies are quantized with values $f_n=nc/L_{rt}$, where $n$ is an integer, $c$ is the speed of light, and $L_{rt}$ is the roundtrip length of the cavity.  The output of a laser is highly peaked around these resonant frequencies, with the width of the peak dependent upon the reflectivity of the mirrors, $R$.  The exact spectrum of a laser's output is also affected by other factors, but one can zero in on the BIQ effect by looking at the transmission through an \textit{empty} (Fabry-P\'erot) cavity, as a function of different frequencies.  Figure 1 shows this transmission fraction in the neighborhood of two of the resonant frequencies (say, $f_n$ and $f_{n+1}$), for three different values of $R$.  

\begin{figure}[htb]
\centerline{\includegraphics[width=.45\textwidth]{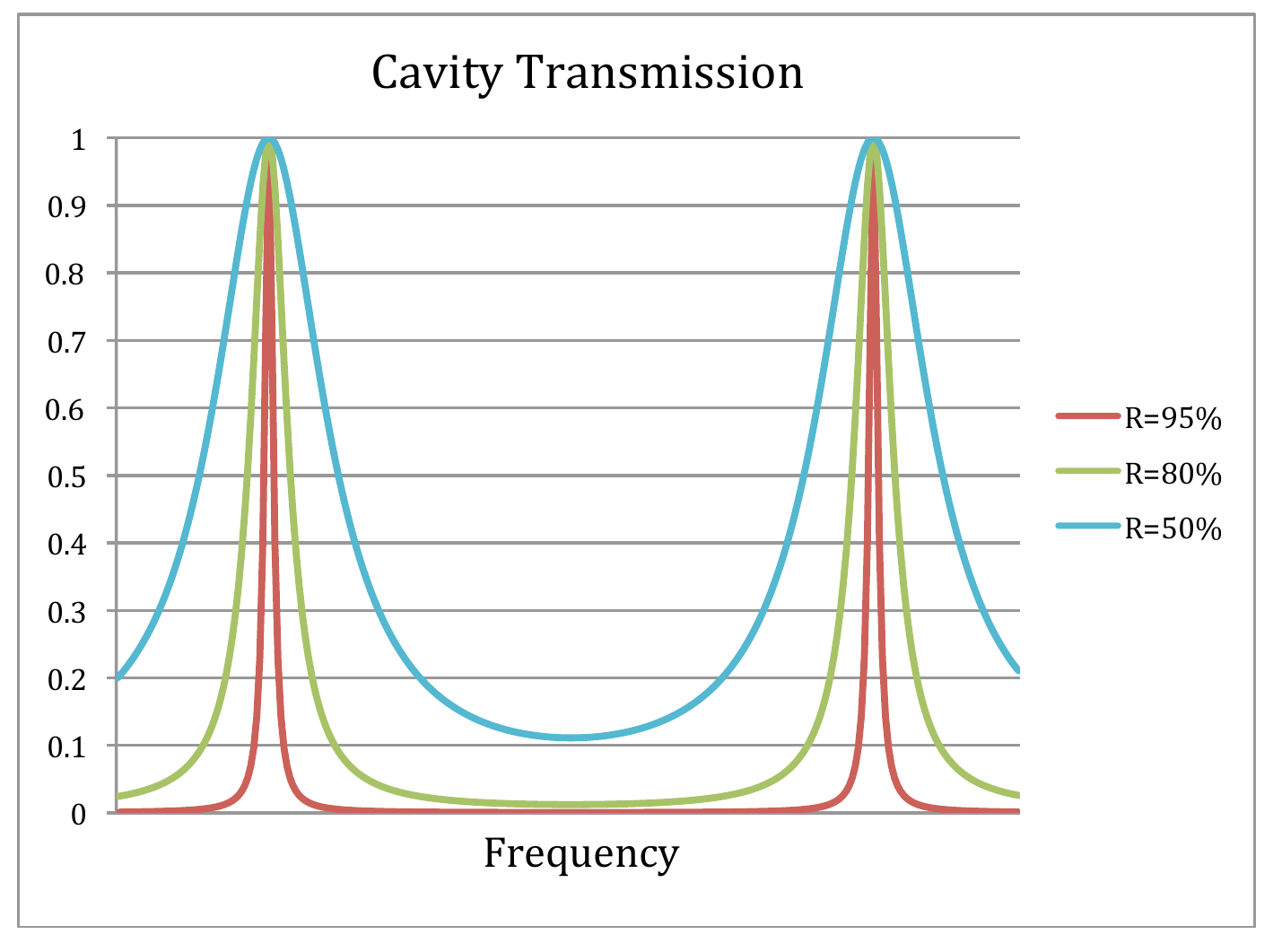}}
\caption{The transmission through a Fabry-P\'erot cavity, as a function of frequency, for mirror reflectivities $R$.}
\end{figure}

As the quality of the cavity increases (or as $R$ approaches 100\%), the resonant modes become closer to being perfectly discrete.  But it's also worth noticing that as the mirrors become less and less of a constraint on the light, the discreteness melts away.  These well-known results can hardly be stressed enough: boundary constraints on continuous systems lead to an emergent discreteness, \textit{with a precision directly linked to how strongly that system is constrained}.  

An analogy to the measurement problem is now within reach.  One of the mysteries of the measurement problem is discussed in the previous section: Stricter measurements produce stricter discreteness.  If there is a connection between the precision of a measurement and the reflectivity of a laser cavity mirror, we would have a beautiful example of how such a variable-discreteness might naturally emerge from basic principles.

But disanalogies seem to abound -- both in terms of the type of discreteness in BIQ vs. QT, but also due to the apparent difference between a mirror and a measurement.  The former issue will be addressed at the end of Section V, and the latter issue in Section IV.  Leading into this next section, note that a mirror is a constraint on the electromagnetic field that is imposed at a particular location (spanning a duration of time).  Conversely, a quantum measurement is traditionally imposed at a particular time (spanning a region of space), and usually isn't viewed as a constraint on the system being measured.  So the analogy appears to be broken -- at least if you take the standard QT-view of measurements.  

But QT comes up short when it comes to defining what exactly is meant by a measurement.  Indeed, some formulations of the measurement problem come down to precisely this unanswered question.  Fortunately, we have another pillar to lean on: General Relativity.  And in GR's framework, as we're about to see, mirrors and measurements are two sides of the same coin.

\section{GR-compatible Measurements}

From the perspective of wanting to unite QT and GR, it makes sense that one would want quantum measurements to be compatible with the GR framework, so that a unified theory would at least be conceivable.  And yet standard quantum measurement theory isn't even consistent with \emph{special} relativity, in that it applies to mathematical structures that mysteriously link distant regions of spacetime.  If we're going to reconcile measurements with GR it's actually easier to just start from scratch, taking the principles of GR and seeing what sort of conclusions naturally follow.  At first pass, the conclusions will seem to contradict ``known'' results from QT, but by the end of the essay I hope to show that if this GR-approach is correct, those ``known'' results are built from possibly-faulty assumptions.

First and foremost, GR is a theory of spacetime.  Not space evolving or flowing through time, but just \emph{spacetime}, a static four-dimensional manifold.  The entities in spacetime also do not ``flow'' or change in any way.  Each spot on the manifold is localized in both space \emph{and} time; a particle would be represented as a motionless line, stretching between past events and the future events.  The all-too-common picture of a particle moving along such a `world-line' is a dangerous error: motion in a spacetime framework is impossible, as motion requires elapsed time, and here time is already spread out into the geometry.  This spacetime viewpoint is often referred to as that of a ``block universe'' \cite{Price}, and it is essential to discuss the GR framework in an objective manner.

In a static spacetime picture, the only thing that might correspond to a physical measurement is the boundary between two spacetime regions.  That's because GR is ``local''; if a measurement is going to be made on events in one block of spacetime, the measurement device can't be in some distant region of spacetime -- it has to be right there, bordering the region in question.\footnote{One could have an intermediate region act as a correlational go-between, but in a local theory there still must be a contiguous path.}

Consider the boundary between any two arbitrary regions of spacetime.  It's a mistake to postulate information moving across from a ``measured region'' to some ``measurement device region''.  (Again, it is a logical error to have an objective flow in a block universe.)  But for any particular physical laws, the boundary will enforce correlations between the two regions.  One example from the previous section is the interface between a region containing an electromagnetic field and a region containing a mirror.  The laws of classical electromagnetism demand that the transverse components of the electric field are zero at the mirror -- and that the mirror can only exist at locations where these fields are zero.  

Still, such correlations don't sound much like measurements.  For one thing, it's known that mere internal correlations between subsystems don't manifest themselves via quantum discreteness (at least on small scales).  And if those correlations have to be further correlated to be considered a measurement, how much is enough?  This issue is deeply connected to another aspect of the quantum measurement problem, where every measurement device is in principle a larger quantum system that in turns needs to be measured by something else.  

It turns out there is a natural solution to this infinite regress -- a solution that is all the more obvious in a block universe.  The key insight is that the boundary of the universe itself includes the Big Bang.  If the Big Bang is the ``ultimate external boundary'', then the portions of the universe that become correlated with the Big Bang are constrained in a rather special way.  Far from being just internal correlations, they are actually constrained by the same boundary data that constrains everything.  And it's this special correlation, as we'll see in a moment, that can best be termed a measurement.

In a purely classical mindset, everything is perfectly correlated with the Big Bang, but in reality we don't live in such a pre-deterministic classical universe.  The boundary data at the Big Bang don't exceed the limits of the Heisenberg uncertainty principle, as witnessed by the structure in the cosmic microwave background.  And the Big Bang is in turn subject to the uncertainty principle when it comes to constraining future events.  For any region A of spacetime, the best possible correlation between the Big Bang and the initial boundary of A is only sufficient to determine 50\% of the classical initial boundary data.  (Say, a particle position but not its velocity.)  One can only constrain A beyond this limit by virtue of additional correlations with the future boundary of A.\footnote{This beats the Heisenberg uncertainty limit, which is permitted in hindsight, for example using two precise positions to deduce the exact velocity that a particle must have had.}

\begin{figure}[htb]
\centerline{\includegraphics[width=.45\textwidth]{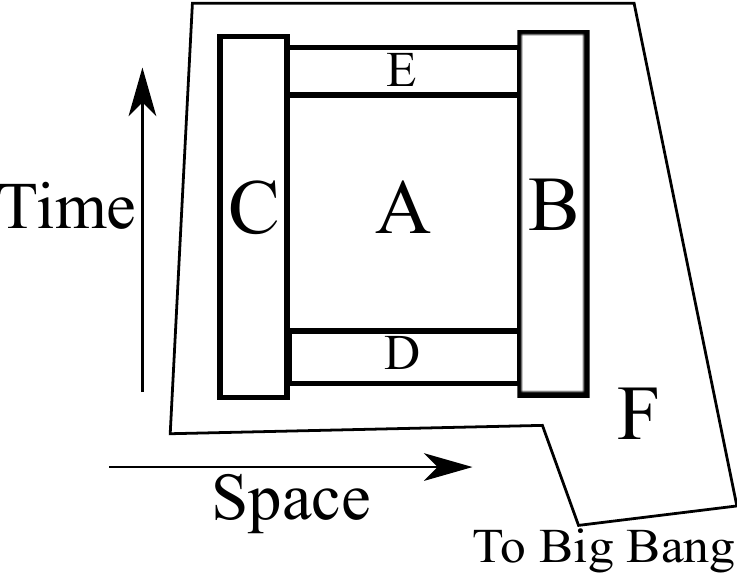}}
\caption{See text.}
\end{figure}

In Figure 2, the quantum system in question is a spacetime region labelled by A.  It is bounded by spatial boundaries (B and C), a preparation region (D), and a future measurement region (E).  Each of these regions become correlated with A by virtue of physical laws; the question is whether we should consider them measurements.  If any of these regions are complex macroscopic structures, it becomes virtually impossible to screen out further correlations with the rest of the universe (F), and the boundary data of A will therefore be linked to the Big Bang data.  This correlation, combined with the given data at the Big Bang, then acts as an objective constraint on A.  In the most constraining case, A can be nearly completely determined, even with only 50\% of the data on each portion of its boundary.\footnote{As in Hamilton's principle; see Section V.}  

But for sufficiently simple regions B,C,D,E, it's possible to screen out the correlation between F and A.  For example, suppose E is an atom that interacts with A on the A/E boundary, correlating some property of A with the energy of the atom.  Few quantum physicists would call this A/E interaction a measurement of A, but if the energy of the atom is later correlated to some macroscopic device in F, then there is a chain of constraints that acts exactly like a direct measurement of A -- and experimentally, A is known to behave accordingly.  The critical point is that a simple system like an atom may never correlate its energy with F, as in so-called quantum eraser experiments.\cite{qeraser}  For example, it may enter some detector that is sensitive to the time of arrival, but not the energy.  This effectively turns the region E into a ``screen'' for any correlations between F and A.  In this case we would say that the A/E interaction was not a measurement, and this conclusion is again borne out by experiment.

And this is exactly the problem with QT's non-block-universe viewpoint.  Because QT operates using only instantaneous descriptions, it can't possibly say whether a particular interaction will \emph{eventually} lead to correlations with the Big Bang -- and so it can never be certain about whether to call a given interaction a measurement or not. The block universe view gets around this problem by simply looking at all of spacetime as one coherent structure.  Here, measurements don't occur at some place and time: they are correlations \emph{through} spacetime.  We just draw the boundaries between regions where we feel it is most useful to do so, but this subjectivity doesn't interfere with the objective nature of the spacetime correlations.

Which brings us to the GR-inspired solution to the ``measurement problem''.  Any arbitrary system is ``measured'' to the extent that the physical parameters bounding that region are correlated with the boundary conditions at the Big Bang.  Screen off such correlations, and the system is unconstrained.  Allow the fullest possible correlations, and the system is highly constrained.  From this perspective, there's really no difference between a careful preparation (D), mirrors on a laser cavity (B and C), or a final measurement (E); each one of these can act like a partial constraint on A.\footnote{The microscopic/macroscopic divide in QT then comes down to complex spacetime regions being typically good at passing through correlations, and simple spacetime regions being typically good screens.}

This proposed link between an ill-defined ``quantum measurement'' and a well-defined external boundary constraint, connects the measurement-induced quantization of Section II with the phenomenon of BIQ in Section III.  Measuring a quantum system becomes directly analogous to constraining a continuous field via an external cavity.  If one turns Figure 2 on its side, viewing system A as trapped between boundary constraints at D and E, it should be clear how this perspective of ``measurement'' points directly to BIQ as the most probable reason for the apparent discreteness of the constrained system.

\section{Reinterpreting Action Principles}

The above picture of measurement certainly appears to be at odds with QT.  For one thing, it couches everything in terms of local interactions in spacetime, which QT has reasons for rejecting.  For another, it relies on Heisenberg's original reading of his uncertainty principle -- where actual parameters are unconstrained by realistic measurements -- rather than a more modern reading where those unconstrained parameters can't even be said to properly exist. 

But \emph{of course} a GR-based measurement theory is at odds with QT; GR is at odds with QT in the first place.  The goal is not to try to find some non-existent middle ground, but rather to see whether there is some new perspective of the GR framework that can cast light on the mysterious formalism of QT.  Such a goal will also require a mathematical framework if the new perspective is to be made precise.

Remarkably, such a framework already exists and is under widespread use.  I'm referring to action principles,\footnote{Specifically Lagrangian, not Hamiltonian, techniques in analytical field theory.} a mathematical tool used by both GR and QT, generally following crucial mathematical steps widely assumed to have no physical significance.  In GR it works something like this:  One calculates a quantity called the Einstein-Hilbert action $S$, for all possible spacetime geometries (metrics), subject to a boundary constraint where certain properties of the metric are fixed on the spacetime boundary.\cite{BCinGR}  (Such a boundary necessarily includes the future edge, as the boundary on A in Figure 2.)  The Einstein equation then results from Hamilton's principle of action extremization $\delta S=0$.  The action is used somewhat differently in quantum field theory, but it's still built upon these classical foundations.

The only new concept needed to pull all this together is to interpret the mathematical ``fixing'' of the spacetime boundary in classical action principles as a literal, physical external constraint on a system -- just as quantum measurement was described in the previous section.  In Figure 2, this would correspond to the regions B,C,D, and E acting as correlation-pathways from the Big Bang to A's boundary, physically constraining A's boundary parameters so that Hamilton's principle can be successfully applied.  If such constraints are not present, then one must necessarily expand the region on which to apply Hamilton's principle.  Therefore, while some systems could be considered in isolation, other systems would require a bigger picture (as in the case of entangled particles \cite{EPW,WMP}).

The reason this approach is so promising is that it piggybacks on the successful use of action principles by GR and QT, while offering something different by linking this mathematics to actual measurements.  In this new perspective, ``fixing the boundary'' shifts from an abstract mathematical procedure to the objective domain of physical interactions.

The most striking consequence is that a quantum system becomes (in part) constrained by events in its own future.  In Fig. 2, the notion of region E constraining region A seems counter-intuitive, but nevertheless this is precisely what the mathematics of Hamilton's principle literally implies.  In exchange for our giving up our time-asymmetric intuitions, we get straightforward answers to deep questions.  The source of the apparent non-locality in QT would be the counter-intuitive dependence of past parameters in A on future measurement settings in E.\footnote{This voids the premises behind Bell's inequality, and therefore takes away the evidence for nonlocality. \cite{Price,EPW,WMP}}  For classical fields, the fact that only half of the boundary data is fixed when using Hamilton's principle starts looking a lot like Heisenberg's original interpretation of measurement uncertainty.\footnote{In this view, the second-order Klein Gordon Equation becomes more fundamental than the first-order Schrodinger equation. \cite{Wharton1}}  

But most importantly, the emergence of quantization upon measurement -- a quantization that is automatically more precise for more constraining measurements -- follows from the exact same logic that explains BIQ in spatial boundary problems.  The only difference is that now the boundary constraints are at two different times, which means that future measurement constraints somehow influence the allowed modes in the past.  Again, if we can get past our instinctive distaste of such a picture, the evident connection between measurement and discreteness falls together quite nicely.  

The final question is whether the BIQ-style discreteness implied by such a picture quantitatively matches the actual discreteness observed in experiments.  Answering this in full will require additional research.  Although it's not obvious how particle-discreteness would arise, there are some reasonable arguments that classical fields constrained by both spatial and temporal boundaries would be forced into particle-like modes via the joint combination of the two constraints.\cite{Wharton2}  It is also unclear how to implement weak measurements, but the qualitative analogy with the reflectivity of a laser cavity mirror is promising.

One recent result stands out as the most successful match between BIQ and QT.  It turns out that there is a simple extension of Hamilton's principle -- demanding that all boundary data lead to an extremizable action -- which provides a strong constraint on measurement outcomes.\cite{Wharton3}  This extension leads to a nearly-discrete set of outcomes from a continuous, classical field that matches up with something called Bohr-Sommerfeld quantization from the old, pre-Schr\"odinger QT.  In particular, the results explicitly show that any angular momentum measurement of a non-relativistic classical Klein-Gordon field must result in very nearly an integer multiple of $\hbar/2$, or else the resulting action would not be extremizable.  Even if this is not what is actually happening when we measure spin, it is a beautiful example of how nearly-discrete outcomes might naturally result from constraining a continuous system.

\section{Summary}

Quantum theory, with its curious blend of continuous equations and discrete measurement theory, is our best window into determining which of these is emergent and which is more fundamental.  I have argued that the discreteness is emergent, based on a number of curious features.  One clue is that discreteness only emerges when certain external measurements are performed.  A bigger clue is that systems are exactly as discrete as required by the precision of the measurement, for their subsequent behavior can only be explained if they are truly continuous below the measured threshold.

Such behavior is exactly what is observed when continuous systems are constrained by spatial boundary conditions (\textit{e.g.} laser cavities).  Not only do discrete modes emerge in such systems, but the precision of the boundary is directly linked to the precision of the discreteness.  It is therefore notable that there is a conceptual link between measurements and boundary constraints, at least when the analysis is restricted to the block universe of general relativity.  In such a block universe, there is no fundamental difference between a system constrained between two spatial boundaries and a system constrained between two consecutive measurements.  Boundary-induced quantization effects are thus not only allowed, they are \emph{expected}.

While the implications of such a picture are counter to our causal intuition, they are nevertheless bolstered by a powerful branch of mathematics (action principles) crucial to both general relativity and quantum field theory.  Using this mathematics as a starting point, the measurement-as-boundary idea has recently led to a natural extension of Hamilton's principle, demonstrating an example of emergent discreteness that matches actual experiments.\cite{Wharton3} 

Although this essay cannot address whether the continuous aspects of quantum theory and general relativity may in turn be emergent from a deeper, discrete reality, the above conclusions argue against the best reasons for suspecting such a digital substructure in the first place.  Quantum theory is not built upon discrete mathematics, but merely dips into it during a problematic and ill-defined measurement process.  At the very least, the ``measurement problem'' should give one pause when drawing digital conclusions from quantum theory.  As for the best, there is promise that we can \emph{solve} the measurement problem by framing it in our continuous block universe.

\end{document}